\documentclass[11pt,a4]{article}
\usepackage{full page}
\usepackage{times}

\usepackage[latin1] {inputenc}
\usepackage{amsmath,amsfonts,amssymb}
\usepackage[dvips]{graphicx}
\usepackage{psfrag}
\usepackage{multicol}

\usepackage{authblk}

\usepackage{color}
\usepackage{ulem}
\newtheorem{definition}{Definition}
\newtheorem{lemma}{Lemma}

\newtheorem{theorem}{Theorem}

\newtheorem{obs}{Observation}

\newcommand{\Prob}[1]{\mathbf{P} \left( #1 \right)}
\newcommand{\Expec}[1]{\mathbf{E} \left[ #1 \right]}

\newcommand{\proof}{\noindent\textit{Proof. }}
\newcommand{\qed}{\hspace{\stretch{1}$\square$}}

\newcommand{\ErdRen}{Erd\H{o}s-R\'{e}nyi }

\newcommand{\grad}{{\mathrm{deg}}}

\newcommand{\edges}{\tbinom{[n]}{2}}

\newcommand{\Expe}[1]{\mathbf{E}\left[ #1 \right]}
\newcommand{\sS}{\mathcal{S}}

\newcommand{\sG}{\mathcal{G}}

\newcommand{\flooding}{\texttt{Flooding} }
\newcommand{\push}{\texttt{Push} }

\newcounter{smallitemizec}
\newenvironment{smallitemize}
{   \setcounter{smallitemizec}{0}
    \vspace{-2ex}
  \begin{list}{$\bullet$}
    {\usecounter{smallitemizec}
      \setlength{\parsep}{0pt}
      \setlength{\itemsep}{0pt}}
    }{ \end{list} 
   \vspace{-2ex}
}

\title{Rumor Spreading in Random Evolving Graphs}
 
 \date{}

\author[1]{Andrea Clementi}
\author[2]{Pierluigi Crescenzi}
\author[3]{Carola Doerr\thanks{Supported by a Feodor Lynen postdoctoral research fellowship of the Alexander von Humboldt Foundation and by the Agence Nationale de la Recherche under the project ANR-09-JCJC-0067-01.}}
\author[4]{Pierre Fraigniaud\thanks{Additional support from the ANR project DISPLEXITY, and from the INRIA project GANG.}}
\author[5]{Marco Isopi}
\author[5]{Alessandro Panconesi}
\author[5]{Francesco Pasquale}
\author[5]{Riccardo Silvestri}

 \affil[1]{ Universit\`a Tor Vergata di Roma}
 \affil[2]{ Universit\`a di Firenze} 
 \affil[3]{ Universit\'e Paris Diderot and Max Planck Institute Saarbr\"ucken} 
 \affil[4]{ CNRS and Universit\'e Paris Diderot,  }
 \affil[5]{ Sapienza Universit\`a di Roma} 

\begin{document}

\maketitle

\begin{abstract}

Randomized gossip is one of the most popular way of disseminating information in large scale networks. This method is appreciated for its simplicity, robustness, and efficiency.  In the \push protocol, every  informed node selects, at every time step (a.k.a. round), one of its neighboring node uniformly at random and forwards the information to this node. This protocol is known to complete information spreading   in $O(\log n)$ time steps  with high probability (w.h.p.) in several 
families of  $n$-node \emph{static} networks.  The \push protocol has also been empirically shown to perform well in practice, and, specifically, to be robust against  dynamic topological changes.

 In this paper, we aim at analyzing the  \push protocol in \emph{dynamic} networks. We  consider the \emph{edge-Markovian} evolving graph model which captures natural temporal dependencies between the structure of the network at time $t$, and the one at time $t+1$. Precisely, a non-edge appears with probability $p$, while an existing edge dies with probability $q$.  In order to fit with real-world traces, we mostly concentrate our study  on the case where $p=\Omega(\frac{1}{n})$ and $q$ is constant. We prove that, in this realistic scenario, the  \push protocol does perform well, completing information spreading in $O(\log n)$ time steps  w.h.p. Note that this performance holds even when the network is, w.h.p., disconnected at every time step (e.g., when $p\ll \frac{\log n}{n}$). Our result provides the first formal argument demonstrating the robustness of the \push protocol against network changes.  We also address other ranges of parameters $p$ and $q$ (e.g., $p+q=1$ with arbitrary $p$ and $q$, and $p=\frac{1}{n}$ with arbitrary $q$). Although they do not precisely fit with the measures  performed on real-world traces, they can be of independent interest for other settings. The results in these cases confirm the positive impact of dynamism.

\end{abstract}

\thispagestyle{empty}
\newpage

\section{Introduction} \label{sec:intro}

\subsection{Context and Objective}

\textit{Rumor spreading} is a well-known gossip-based distributed algorithm for disseminating information in large networks. According to the synchronous \push version of this algorithm, an arbitrary source node is initially informed, and, at each time step (a.k.a. round), an informed node $u$ chooses one of its neighbors $v$ uniformly at random, and this node becomes informed at the next time step. 

Rumor spreading (originally called \textit{rumor mongering}) was first introduced by~\cite{DGHILSSST87}, in the context of replicated databases, as a solution to the problem of distributing updates and driving replicas towards consistency. Successively, it has been proposed in several other application areas, such as failure detection in distributed systems~\cite{VMH98}, peer-sampling~\cite{JVGKS07}, adaptive machine discovery~\cite{HTL99}, and distributed averaging in sensor networks~\cite{BABD05} (for a nice survey of gossip-based algorithm applications, see also~\cite{KS07}). Apart from its applications, rumor spreading has also been deeply analyzed from a theoretical and mathematical point of view. Indeed, as already observed in~\cite{DGHILSSST87}, rumor spreading is just an example of an epidemic process: hence, its analysis ``benefits greatly  from the existing mathematical theory of epidemiology'' (even if its application in the field of distributed systems has almost opposite goals). In particular, the \textit{completion time} of rumor spreading, that is, the number of steps required in order to have all nodes informed with high probability\footnote{An event holds with high probability  if it holds with probability at least $1 - 1/n^c$ for some constant $c>0$.} (w.h.p.), has been investigated in the case of several different network topologies, such as complete graphs~\cite{FG85,Pi87,KSSV00}, hypercubes~\cite{FPRU90}, random graphs~\cite{FPRU90,FHP10,FP10},  preferential attachment graphs~\cite{CLP09,DFF11}, 
 and  some  power-law degree graphs~\cite{FountoulakisPS12}. Besides obtaining bounds on the completion time of  rumor spreading, most of these works also derive deep connections between the completion time itself and some classic measures of graph spectral theory, such as, for example, the \textit{conductance} of a graph (as far as we know, the most recent results of this kind are the ones presented in~\cite{CLP10,CLP10b,G11}) or its \textit{vertex expansion} (see~\cite{SS11, GS12}). 
 
It is important to observe that the techniques and the arguments adopted in these studies strongly rely on the fact that the underlying graph is \textit{static} and does not change over time. For instance, most of these analyses exploit the  crucial fact that the degree of every node (no matter whether  this is a random variable or a deterministic value)  never changes during the entire execution of the rumor spreading algorithm. It is then natural to ask ourselves what is the speed of rumor spreading in the case of \textit{dynamic} networks, where nodes and edges can appear and disappear over time (several emerging networking technologies such as ad hoc wireless, sensor, mobile networks, and peer-to-peer networks are indeed inherently dynamic). 

In order to investigate the behavior of distributed protocols in the case of dynamic networks, the concept of evolving graph has been introduced in the literature. An \textit{evolving graph} is a sequence of graphs $(G_t)_{t\geq 0}$ where $t\in \mathbb{N}$ (to indicate that we consider the graph \emph{snapshots} at   discrete time steps $t$, although it may evolve in a continuous manner) with the same set of $n$ nodes.\footnote{As far as we know, this definition has been formally introduced for the first time in~\cite{F02}.} This concept is  general enough for allowing us to model basically any kind of network evolution, ranging from \textit{adversarial} evolving graphs (see, for example,~\cite{CMPS07,KLO10}) to  \textit{random} evolving graphs (see, for example,~\cite{B2001}). 

Indeed, although only the edges are subject to changes, a node whose all incident edges are not present at a given step $t$ can be considered as having left the network at time $t$, where the network is viewed as the giant component of $G_t$. Hence, the concept of evolving graph also captures some essence of the node dynamics. In the case of \emph{random} evolving graphs, at each time step, the graph $G_t$ is chosen randomly according to some probability distribution over a specified family of graphs. One very well-known and deeply studied example of such a family is the set $\sG_{n,p}$ of \textit{Erd\H os-R\'enyi} random graphs~\cite{AKL08,ER59,G59}. In the evolving graph setting, at every time step $t$, each possible edge exists with probability $p$ (independently of the previous  graphs $G_{t'}$, $t'<t$,  and independently of  the other edges in $G_t$). 

Random evolving graphs can exhibit communication properties which are much stronger than static networks having the same expected edge density (for a recent survey on computing over dynamic networks, see~\cite{KO11}).  This has been proved in the case of the simplest communication protocol that implements the broadcast operation, that is, the \flooding protocol (a.k.a. \textit{broadcasting protocol}), according to which a source node is initially informed, and, whenever  an uninformed node has an informed neighbor, it becomes informed itself at the next time step.  It has been shown~\cite{BCF11,CMMPS08,CMPS09} that the \flooding completion time may be  very fast (typically poly-logarithmic in the number of nodes) even when the network topology is, w.h.p., sparse, or even highly disconnected at every time step. Therefore, such previous results provide analytical evidences of the fact that random network dynamics not only do not hurt, but can actually help data communication, which is of the utmost importance in several contexts, such as, e.g., delay-tolerant networking~\cite{V11,WCA11}.  

The same observation has been made when the model includes some sort of \textit{temporal} dependency, as it is in the case of the random \textit{edge-Markovian} model. According to this model, the evolving graph starts with an arbitrary initial graph $G_0$, and, at every time step $t$,
\smallskip
\begin{smallitemize}
\item if an edge does not exist in $G_t$, then it will appear in the next graph $G_{t+1}$ with probability $p$, and 
\item if an edge exists in $G_t$, then it will disappear in the next graph $G_{t+1}$ with probability $q$. 
\end{smallitemize}
\smallskip
For every initial graph $G_0$, an edge-Markovian evolving graph will eventually converge to a (random) graph in $\sG_{n,\tilde{p}}$ with stationary edge-probability $\tilde{p}=\frac{p}{p+q}$. 
However, there is a Markovian dependence between graphs at two consecutive time steps, hence, given $G_t$, the next graph $G_{t+1}$ is not necessarily a random graph in $\sG_{n,\tilde{p}}$. Interestingly enough, the edge-Markovian model has been recently subject to experimental validations, in the context of sparse opportunistic mobile networks~\cite{WCA11}, and of dynamic  peer-to-peer systems~\cite{V11}. These validations demonstrate a good fitting of the model with some real-world data traces. The completion time of the  \flooding protocol has been recently analyzed in this model, for all possible values of $\tilde{p}$ (see~\cite{BCF11,CMPS09}). A variant of the model, in which the ``birth'' and ``death'' probabilities $p$ and $q$ depend not only on the number of nodes but also on some sort of distance between the nodes, has been investigated in~\cite{GH10}.  

The \flooding protocol however generates high message complexity. Moreover, although its completion time is an interesting analog for  dynamic graphs of the diameter for static graphs, it is not reflecting the kinds of gossip protocols mentioned at the beginning of this introduction, used for practical applications. Hence the main objective of this paper is to analyze the more practical \push protocol, in edge-Markovian evolving graphs. 

\subsection{Framework} 

We focus our attention on dynamic network topologies yielded by the edge-Markovian evolving graphs for parameters $p$ ($b$irth) and $q$ ($d$eath) that correspond to a good fitting with real-world data traces, as observed in~\cite{V11,WCA11}. These traces describe  networks with relatively high dynamics, for which the death probability $q$ is at least one order of magnitude greater than the birth probability $p$. In order to set parameters $p$ and $q$ fitting with these observations, let us consider the expected number of edges $\bar{m}$, and the expected node-degree $\bar{d}$ at the stationary regime, governed by $\tilde{p}=\frac{p}{p+q}$.   We have 
$
\bar{m} = \frac{p}{p+q} \binom{n}{2},
$
and
$
\bar{d} = \frac{2\bar{m}}{n} = (n-1) \frac{p}{p+q}.
$
Thus, at the stationary regime, the expected number of edges $\nu$ that switch their state (from non existing to existing, or vice versa) in one time step satisfies
\[\textstyle
\nu = \bar{m} q + (  \binom{n}{2} - \bar{m} ) p = \frac{n (n-1)}{2}\left(\frac{pq}{p+q}+\left(1-\frac{p}{p+q}\right)p\right) = n (n-1) \frac{pq}{p+q} = n q \bar{d}.
\]
Hence, in order to fit with the high dynamics observed in real-world data traces, we set $q$ constant, so that a constant fraction of the edges disappear at every step, while a fraction $p$ of the non-existing edges appear. We consider an arbitrary range for $p$, with the unique assumption that $p\ge \frac{1}{n}$. (For smaller $p$'s, the completion time of any communication protocol is subject to the expected time $\frac{1}{np}\gg 1$ required for a node to acquire just one link connected to another node). To sum up, we essentially focus on the following range of parameters:  
\begin{equation}\label{parranges}
    \frac {1} n \ \leqslant p \  < 1  \  \mbox{ and } \  q = \Omega(1).
\end{equation} 
This range includes network  topologies for a wide interval of expected edge density (from very sparse and disconnected graphs, to almost-complete ones), and with an expected number of switching edges per time step equal to some constant fraction of the expected total number of edges. Other ranges are also analyzed in the paper (e.g., $p+q=1$ with arbitrary $p$ and $q$, and $p=\frac{1}{n}$ with arbitrary $q$), but the range in Eq.~\eqref{parranges} appears to be the most realistic one, according to the current measurements on dynamic networks.  

\paragraph{Remark.} 
It is worth noticing that analyzing the  \push protocol in edge-Markovian graphs is not only subject to temporal dependencies, but also to \emph{spatial} dependencies. This makes the analysis of the  \push protocol more challenging. This holds even in the simpler random evolving graph model, i.e., the sequence of independent random graphs  $G_t\in \sG_{n,p}$. Indeed, even if this case does not include temporal dependencies, the \push protocol introduces spatial dependences that has to be carefully handled. To see why, consider a time step of the \push protocol, where we have $k$ informed nodes, and let us try  to evaluate how many new informed nodes there will be in the next time step. Given an informed node $u$, let $\delta(u)$ be the neighboring node selected by $u$ according to the \push protocol (i.e., $\delta(u)$ is chosen uniformly at random among the current neighbors of $u$). By conditioning on the degree of $u$, it is not hard to calculate the probability that $\delta(u) = v$,  for any non informed node $v$. However, the events ``$\delta(u_1) = v_1$'' and ``$\delta(u_2) = v_2$'' are not necessarily independent. Indeed, the event ``$\delta(u_1) = v_1$'' decreases   the probability of the existence of an edge between $u_1$ and $u_2$, and so it affects the value of the random variable $\delta(u_2)$. This  positive  dependency prevents us from using the classical methods for analyzing the \push protocol in static graphs, or makes the use of these methods far more complex.

\subsection{Our results} 

For the  parameter range in Eq.~\eqref{parranges}, we show that, w.h.p., starting from any $n$-node graph $G_0$, the \push protocol informs all $n$ nodes in $\Theta(\log n)$ time steps.   Hence, in particular, even if the graph $G_t$ is w.h.p. disconnected at every time step (this is the case
 for $p\ll \frac{\log n}{n}$), the completion time of the \push protocol is as small as it could be (the \push protocol cannot perform faster than $\Omega(\log n)$ steps in any static or dynamic graph since the number of informed nodes can at most double at every step). It is also interesting to compare the performances of the \push protocol with the one of \flooding. The known lower bound for \flooding on edge-Markovian graphs~\cite{CMPS09} (which  is clearly a lower bound for \push, too) demonstrates that   for $p = \Theta(1/n)$,  the two protocols have the same asymptotic completion time. Moreover it is clear that, for $p=\Omega(1/n)$, the completion-time slowdown factor of the \push\ protocol  
 is at most      logarithmic.
 This  property is a remarkable one, since the  expected 
 number of exchanged  messages per node in \push may  be  exponentially smaller than the one in \flooding 
 (for instance, consider the case $p=\Theta(1/\sqrt n)$ which corresponds to an expected node degree $\Theta(\sqrt n)$). 

We also address other ranges of parameters $p$ and $q$. Although they do not precisely fit with the measures in~\cite{V11,WCA11}, they can be of independent interest for other settings. One such case is the sequence of independent $\sG_{n,p}$ graphs, that is, the case where $p+q=1$. Actually, the analysis of this  special  case will allow us to  focus  on  the first important probabilistic issue that needs to be solved: spatial dependencies. Indeed, even in this case, as already mentioned, the \push protocol induces a positive correlation among some crucial events that determine the number of new informed nodes at the next time step. This holds despite the fact that every edge is set independently from the others.   For a sequence of independent $\sG_{n,p}$ graphs, we prove that for every $p$ (i.e., also for $p = o(\frac 1n)$) and $q=1-p$ the completion time of the \push protocol is, w.h.p., $\mathcal{O}(\log n / (\hat p n))$, where $\hat{p} = \min\{p,\, 1/n \}$. By comparing the lower bound for \flooding in \cite{CMPS09}, it turns out that this  bound is tight,  even for very sparse graphs. 

Finally, we  show that the logarithmic bound for the \push protocol holds for   more   ``static'' network topologies as well, e.g., for the range $p \ = \ \frac {c} n$ where $c>0$ is a constant, and $q$ is arbitrary. This parameter range includes edge-Markovian graphs with a small expected number of switching edges (this happens when $q = o(1)$). In this case, too, \push completes, w.h.p., in $O(\log n)$ rounds. This gives yet another evidence that dynamism helps. 

\paragraph{Structure of the paper.} In Section~\ref{sec:pre}, we give the  terminology and the preliminary  definitions that will be used throughout the paper. In Section \ref{sec:indgnp}, we   consider the independent dynamic Erd\H os-R\'enyi   graphs, while Section \ref{sec:edgemeg} provides the analysis of the \push protocol in the the case of the edge-Markovian evolving graph model. In Section \ref{sec:conc}, finally, we summarize our results and present their extension to the case of more ``static'' network topologies.

\section{Preliminaries} \label{sec:pre}

The number of vertices in the graph will always be denoted by $n$. We abbreviate $[n]:=\{1, \ldots, n\}$ and $\edges:=\{\{i,j\} \mid i,j \in [n]\}$. For any subset $E \subseteq \edges$ and  any two   subsets $A,B \subseteq [n]$, define
\[
E(A) = \{  \mbox{ edges of  $E$ incident to $A$   } \} \; \text{ and } \; E(A,B) =  \{   \{u,v\} \in E \ | \ u \in A , v \in B  \}.
\]
We   consider the edge-Markovian evolving graph model  $\sG(n,p,q;E_0)$ where $E_0$ is the starting set of edges.

\noindent
The \push Protocol over $\sG(n,p,q;E_0)$ can be represented as a random process over the set $\sS$ of all possible pairs $(E,I)$ where $E$ is a subset of edges and $I$ is a subset of nodes.
In particular, the combined Markov process works as follows

\[
\ldots \rightarrow \ (E_t,I_t) 
\stackrel{ \mbox{ {\tiny edge-Markovian} }}{ \longrightarrow } 
(E_{t+1},I_{t})
\stackrel{ \mbox{{\tiny  \push protocol}}}{ \longrightarrow }  
(E_{t+1},I_{t+1})   \stackrel{ \mbox{ {\tiny edge-Markovian} }}{ \longrightarrow }  \ldots
\]
where $E_t$ and $I_t$ represent the set of existing edges and the set of informed nodes at time $t$, respectively. All events, probabilities and random variables are defined over the above random process.
Given a graph $G = ([n],E)$, a node $v \in [n]$, and a subset of nodes $A \subseteq [n]$ we define  $\grad_G(v,A) = $ $ |\{(v,a) \in E \ | \  a \in A \}|$.
 When we have a sequence of graphs $\{ G_t = ([n],E_t) \,:\, t \in \mathbb{N} \}$ we write $\grad_t(v,A)$ instead of $\grad_{G_t}(v,A)$. Given a graph $G$ and an informed node $u \in I$, we define $\delta_G(u)$ as the random variable indicating the node selected by $u$ in graph $G$ according to the \push protocol. When $G$ and/or $t$ are clear from the context, they will be omitted.

\section{Warm up: the time-independent case} \label{sec:indgnp}

In this section we analyze the special case of a sequence of independent $G_{n,p}$   (observe that a sequence of independent $G_{n,p}$   is edge-Markovian with $q = 1-p$). We show that the completion time of the \push protocol is $\mathcal{O}(\log n / (\hat{p} n))$  w.h.p., where $\hat{p} = \min\{p,\, 1/n \}$. In Theorem~\ref{theorem:indepover} we prove the result for $p \geqslant 1/n$ and in Theorem~\ref{theorem:indepunder} for $p \leqslant 1/n$. From the lower bound on the flooding time for edge-Markovina graphs \cite{CMPS09}, it turns out that our bound is optimal.

As mentioned in the introduction, even though in this case there is no time-dependency in the sequence of graphs, the \push protocol introduces a kind of dependence that has to be carefully handled. 
 The key challenge is to evaluate the probability that $v$ receives the information from at least one of the informed nodes; i.e.,
$1 - \Prob{ \cap_{u \in I} \{\delta(u) \neq v\} }$.
We consider the \push operation on a \emph{modified} random graph where we prove that the above events become independent and the number of new informed nodes in the original random graph is at least as large as in the modified version.

\begin{definition}[$(I,b)$-modified graph]\label{def:modifiedgraph}
Let $G = ([n],E)$ be a graph, let $I \subseteq [n]$ be a set of nodes, and let $b \in [n]$ be a positive integer. The $(I,b)$-modified $G$ is the graph $H=([n] \cup \{v_1, \dots, v_b \})$, where $\{v_1, \dots, v_b\}$ is a set of extra \emph{virtual} nodes, obtained from $G$ by the following operations:
1. For every node $u \in I$ with $\grad_G(u) > b$, remove all edges incident to $u$;
2.  For every node $u \in I$ with $\grad_G(u) \leqslant b$, add all edges $\{ u,v_1\}, \dots, \{u,v_b\}$ between $u$ and the virtual nodes;
3. Remove all edges between any pair of nodes that are both in $I$.
\end{definition}

\noindent
Let $I$ be the set of informed nodes performing a \push operation on a $G_{n,p}$ random graph. 
As previously observed, if $v \in [n]\setminus I$ is a non-informed node, then   the  events $\left\{ \{\delta_G(u) = v \} \,:\, u \in I \right\}$ are not independent, but 
the events $\left\{ \{\delta_H(u) = v \} \,:\, u \in I \right\}$ on the $(I,b)$-modified graph $H$ are independent because of Operation 3 in Definition \ref{def:modifiedgraph}. 

In the next lemma we prove that, if the informed nodes perform a \push operation both in a graph  and in its modified version, then the number of new informed nodes in the original 
graph  is (stochastically) larger than the number of informed nodes in the modified one. We will then apply this result to   $G_{n,p}$ random graphs.

\begin{lemma}[Virtual nodes]\label{lemma:virtualnodes}
Let  $G([n],E)$ be a graph and let $b$ an integer such that $1 \leqslant b \leqslant n$.
Let $I \subseteq [n]$ be a set of nodes performing a \push operation in graphs $G$ and $H$, where $H$ is the $(I,b)$-modified $G$ according to Definition~\ref{def:modifiedgraph}.
Let $X$ and $Y$ be the random variables counting the numbers of new informed nodes in $G$ and $H$ respectively. Then for every $h \in [0,n]$ it holds that
$
\Prob{X \leqslant h} \leqslant \Prob{Y \leqslant h}$.
\end{lemma}
\proof
Consider the following coupling: Let $u \in I$ be an informed node such that $\grad_G(u) \leqslant b$ and let $h$ and $k$   be the number of informed and non-informed neighbors of $u$ respectively. Choose $\delta_H(u)$ u.a.r. among the neighbors of $u$ in $H$. As for $\delta_G(u)$, we do the following:
If $\delta_H(u) \in [n] \setminus I$ then choose $\delta_G(u) = \delta_H(u)$; 
otherwise (i.e., when $\delta_H(u)$ is a virtual node) with probability $1-x$ choose $\delta_G(u)$ u.a.r. among the informed neighbors of $u$ in $G$, and  with probability 
$x$ choose $\delta_G(u)$ u.a.r. among the non-informed ones, where $x = \frac{k(b-h)}{(h+k)b} $.
Every informed node $u$ with $\grad_G(u) > b$ instead performs  a \push operation in $G$ independently.

By construction we have that the set of new (non-virtual) informed nodes in $H$ is a subset of the set of new informed nodes in $G$. Moreover, it is easy to check that,
 for every informed node $u$ in $I$, $\delta_G(u)$ is u.a.r. among neighbors of $u$. 
\qed

\smallskip\noindent
In the next lemma we give a lower bound on the probability that a non-informed node gets informed in the  modified $G_{n,p}$.

\begin{lemma}[The increasing rate of informed nodes]
\label{lemma:incratindip}
Let $I \subseteq [n]$ be the set of informed nodes performing the \push operation in a $G_{n,p}$ random graph and let $X$ be the random variable counting the number of non-informed nodes that get informed after the \push operation. It holds that
$
\Prob{X \geqslant \lambda \cdot \min\{ |I|, \, n-|I| \}} \geqslant \lambda 
$, 
where $\lambda$ is a positive constant.
\end{lemma}
\proof
Let $I$ be the set of currently informed nodes, let $G=([n],E)$ be the random graph at the next time step and let $H$ be its $(I, 3np)$-modified version. Now we show that the number of nodes that gets informed in $H$ is at least $\lambda \cdot \min\{ |I|, n-|I| \}$ with probability at least $\lambda$, for a suitable constant $\lambda$.

\noindent
Let $u \in I$ be an informed node and let $v \in [n]\setminus I$ be a non-informed one. Observe that by the definition of $H$, $u$ cannot choose $v$ in $H$ if the edge $\{ u,v \} \notin E$ or if the degree of $u$ in $G$ is larger than $3 n p$ (see Operation 3 in Definition \ref{def:modifiedgraph}). Thus the probability that node $u$ chooses node $v$ in random graph $H$ according to the \push protocol is  
\begin{equation}\label{eq:probinH}
\Prob{\delta_{H}(u) = v} = \Prob{\delta_{H}(u) = v \;|\; \{ u,v \} \in E \, \wedge \, \grad_G(u) \leqslant 3 n p } \Prob{\{ u,v \} \in G \,\wedge \, \grad_G(u) \leqslant 3 n p}.
\end{equation}

\noindent
If $\grad_G(u) \leqslant 3 n p$ then  node $u$ in $H$ has exactly $3np$ virtual neighbors plus at most other $3np$ non-informed neighbors.
It follows that 
\begin{equation} \label{eq:prH1}
\Prob{\delta_{H}(u) = v \,|\, \{ u,v \} \in E \, \wedge \, \grad_G(u) \leqslant 3 n p } \geqslant 1/(6np).
\end{equation} 

\noindent
We also  have that  
\begin{eqnarray*}
\Prob{\{ u,v \} \in E, \, \grad_G(u) \leqslant 3 n p} 
& = & \Prob{\{ u,v \} \in E} \Prob{\grad_G(u) \leqslant 3 n p \; | \; \{ u,v \} \in E}  \nonumber \\
& = & p \cdot \Prob{\grad_G(u) \leqslant 3 n p \; | \; \{ u,v \} \in E}. 
\end{eqnarray*}
Since 
$\Expec{\grad_{G}(u) \,|\, \{ u,v \} \in E} \leqslant np + 1 \ \mbox{ with }  \  np \geqslant  1$,  from the Chernoff bound we can choose
a  positive constant $c$ and then a positive constant $\beta <1$ such that
\begin{equation} \label{eq:prH3}
\Prob{\grad_{G}(u) > 3 np \,|\, \{ u,v \} \in E} \leqslant \Prob{\grad_{G}(u) > 2 np + 1 \,|\, \{ u,v \} \in E} \leqslant e^{- c np} = \beta < 1.
\end{equation}
  By replacing Eq.s~\ref{eq:prH1} and~\ref{eq:prH3} into Eq.~\ref{eq:probinH} we get 
  $ \Prob{\delta_{H}(u) =  v} \ \geqslant    \ \frac{\alpha} n$,  for some  constant $ \alpha >0$. 
  
  \noindent 
  Since  the  events $\{ \{\delta_H(u) = v \}, v \in I \}$ are independent, the probability that node $v$ is not informed in $H$ is thus
$$
\Prob{\cap_{u \in I} \delta_{H}(u) \neq v}
\leqslant \left( 1 - \alpha / n \right)^{|I|} 
\leqslant e^{- \alpha |I| / n}.
$$
Let $Y$ be the random variable counting the number of new informed nodes in $H$. The expectation of $Y$ is
$$
\Expec{Y} \geqslant (n-|I|)\left( 1 - e^{- \alpha |I| / n} \right) \geqslant (\alpha / 2) (n-|I|)|I| / n.
$$
Hence we get 
$$
\Expec{Y} \geqslant
\left\{
\begin{array}{cl}
(\alpha / 4) |I| & \quad \mbox{ if } |I| \leqslant n/2\,, \\[2mm]
(\alpha / 4) (n-|I|) & \quad \mbox{ if } |I| \geqslant n/2\,.
\end{array}
\right.
$$
Since $Y \leqslant \min \{ |I|,\, n-|I| \}$, from Observation~\ref{obs:expecttoprob} (see Appendix~\ref{app:observations}), it follows that \\
$
\Prob{Y \geqslant (\alpha/8) \cdot \min \{ |I|,\, n-|I| \}} \geqslant \alpha/8$.
Finally we get the thesis by applying  Lemma~\ref{lemma:virtualnodes}.
\qed

\smallskip
\noindent
We can now derive the upper bound on the completion time of the \push protocol on   $G_{n,p}$ random graphs.

\begin{theorem}\label{theorem:indepover}
Let $\mathcal{G} = \{G_t \,:\, t \in \mathbb{N}\}$ be a sequence of independent $G_{n,p}$ with $p \geqslant 1/n$. 
The completion time of the \push protocol over $\mathcal{G}$ is $\mathcal{O}(\log n)$ w.h.p.
\end{theorem}
\proof
Consider a generic time step $t$ of the execution of the \push protocol where $I_t \subseteq [n]$ is the set of informed nodes and $m_t = |I_t|$ is its size.  For any   $t$ such that 
 $m_t \leqslant n/2$,   Lemma~\ref{lemma:incratindip} implies  that 
$\Prob{ m_{t+1} \geqslant (1+\lambda) m_t }  \geqslant \lambda$,    where $ \lambda$    is a positive constant.  
Let us define event
$
\mathcal{E}_t = \{ m_{t} \geqslant (1+\lambda) m_{t-1} \} \vee  \{ m_{t-1} \geqslant n/2 \}
$
and let $Y_t = Y_t((E_1,I_1), \dots, (E_t,I_t))$  be the indicator random variable of that event.  Observe that if $t=\frac{\log n}{\log (1+\lambda)}$ then $(1+\lambda)^t \geqslant n/2$. Hence, if we set
$
T_1 = \frac{2}{\lambda} \frac{\log n}{\log (1+\lambda)}$,
we get 
$$
\Prob{m_{T_1} \leqslant n/2} \leqslant \Prob{\sum_{t=1}^{T_1} Y_t \leqslant (\lambda/2)T_1}\,.
$$

\noindent
The above probability  is  at most as large as the probability that in a sequence of $T_1$ independent coin tosses, each one giving \texttt{head} with probability $\lambda$, we see less than $(\lambda/2)T_1$  \texttt{heads}  (see e.g. Lemma 3.1 in \cite{ABKU99}). A direct application of  the Chernoff bound 
shows that  
 this  probability is smaller than 
 $e^{- (1/4) \lambda T_1} \leqslant n^{-c}$,  for a suitable   constant $c >0$.
We can thus state that, after $\mathcal{O}(\log n)$ time steps, there at   least $n/2$ informed nodes w.h.p.

\noindent
If $m_{T_1} \geqslant n/2$, then, for every $t \geqslant T_1$,     Lemma~\ref{lemma:incratindip}
implies that 
$\Prob{ n -m_{t+1} \leqslant (1-\lambda) (n - m_t)} \ \geqslant \lambda$.
Observe that if $t = \frac{\log n}{\lambda}$ then 
$ (1-\lambda)^t \leqslant 1/n$,  so that for 
$T_2 = \frac{2}{\lambda} \cdot \frac{\log n}{\lambda} + T_1$
the probability that the \push protocol has not completed at time $T_2$ is
$$
\Prob{m_{T_2} < n} 
\leqslant 
\Prob{m_{T_2} < n \,|\, m_{T_1} \geqslant \frac{n}{2}} + \Prob{m_{T_1} < \frac{n}{2}}.
$$
As we argued in the analysis of the spreading till $n/2$, the probability 
$
\Prob{m_{T_2} < n \,|\, m_{T_1} \geqslant \frac{n}{2}}$
is not larger than the probability that in a sequence of $\frac{2}{\lambda} \cdot \frac{\log n}{\lambda}$ independent coin tosses, each one giving  \texttt{head} with probability $\lambda$, there are less than $\frac{\log n}{\lambda}$  \texttt{heads}. Again, by applying  the Chernoff bound, the latter is not larger than $n^{-c}$ for a suitable positive constant $c$.
\qed 

\smallskip
\noindent
In order to prove the bound for $p \leqslant 1/n$, we first show that one single \push operation over the union of a sequence of graphs informs (stochastically) less nodes than the sequence of \push operations performed in every single  graph (this fact  will also be used in Section~\ref{sec:edgemeg} to analyse the edge-MEG).

\begin{lemma}[Time windows]\label{lemma:couplingsequences}
Let $\{ G_t = ([n],E_t) \,:\, t = 1, \dots, T\}$ be a finite sequence of graphs with the same set of nodes $[n]$. Let $I \subseteq [n]$ be the set of informed nodes in the initial graph $G_1$. Suppose that at every time step every informed node performs a \push operation, and let $X$ be the random variable counting the number of informed nodes at time step $T$.
Let 
$
H = ([n],F) \mbox{ be such that }  F = \cup_{t=1}^T E_t
$
and let $Y$ be the random variable counting the number of informed nodes when the nodes in $I$ perform one single \push operation in graph $H$. Then for every $\ell = 0,1, \dots, n$ it holds that
$
\Prob{X \leqslant \ell} \leqslant \Prob{Y \leqslant \ell}.
$
\end{lemma}
\proof
Consider the sequence of graphs $\{H_t = ([n],F_t)\,:\, t=1, \dots, T\}$ where graph $H_t$ is the union of graphs $G_1, \dots, G_t$, i.e. for every $t$ we set $F_t = \bigcup_{i=1}^t E_i$.
We inductively construct one single \push operation in $H \equiv H_T$, building it on the probability space of the \push protocol in $(G_1, \dots, G_T)$, in a way that the set of informed nodes in $H$ is a subset of the set of informed nodes in $G_T$.

For every node $u$ that is informed at the beginning of the process, i.e. $u \in I$, and for every $t = 1, \dots, T$, let $N_t$ be the set of neighbors of $u$ in graph $G_t$, let $d_t = |N_t|$ be its size, let $h_t = |\bigcup_{i=1}^t N_i|$ be the number of neighbors of $u$ in graph $H_t$, and let $\delta_{G_t}(u)$ be the random variable indicating the neighbor chosen by $u$ u.a.r. in $N_t$. Finally, let $\{ C_t \,:\, t = 2,\dots, T \}$ be a sequence of independent Bernoulli random variables with $\Prob{C_t = 1} = d_t / h_t$. Now we recursively define random variables $\delta_{H_1}(u), \dots, \delta_{H_T}(u)$:

\smallskip\noindent
Define $\delta_{H_1}(u) = \delta_{G_1}(u)$. For $t=2,\dots, T$ define
\begin{equation}\label{eq:tstepcoupling}
\delta_{H_t}(u) =
\left\{
\begin{array}{cl}
\delta_{G_t}(u) & \quad \mbox{ if } \delta_{G_t}(u) \in N_t \setminus \left( \bigcup_{i=1}^{t-1} N_i \right) \mbox{ and } C_t = 1 \\[2mm]
\delta_{H_{t-1}}(u) & \quad \mbox{ otherwise }
\end{array}
\right.
\end{equation}
By construction, it holds that $\delta_{H_T}(u) \in \{\delta_{G_1}(u), \dots, \delta_{G_T}(u) \}$, hence the set of informed nodes in $H_T$ is a subset of the set of informed nodes in $G_T$. Now we show that for every $t$ node $u$ chooses one of its neighbors uniformly at random in $H_t$, i.e. for every $v \in \bigcup_{i=1}^{t} N_i$ it holds that $\Prob{\delta_{H_t}(u) = v} = 1/h_t$.

We proceed by induction on $t$. The base of the induction directly follows from the choice $\delta_{H_1}(u) = \delta_{G_1}(u)$. Now assume that for every $v \in \bigcup_{i=1}^{t-1} N_i$ it holds that $\Prob{\delta_{H_{t-1}}(u) = v} = 1/h_{t-1}$ and let $v \in \bigcup_{i=1}^t N_i$. We distinguish two cases:

\smallskip \noindent
- If $v \in N_t \setminus \left( \bigcup_{i=1}^{t-1} N_i \right)$ then, according to (\ref{eq:tstepcoupling}) we have that $\delta_{H_t}(u) = v$ if and only if $\delta_{G_t}(u) = v$ and $C_t = 1$, hence
$$
\Prob{\delta_{H_t}(u) = v} = \Prob{\delta_{G_t}(u) = v \, \wedge \, C_t = 1} = \frac{1}{d_t} \cdot \frac{d_t}{h_t} = \frac{1}{h_t}
$$

\smallskip\noindent
- If $v \in \bigcup_{i=1}^{t-1} N_i$ then we have that $\delta_{H_t}(u) = v$ if and only if $\delta_{H_{t-1}}(u) = v$ and at least one of the two conditions in (\ref{eq:tstepcoupling}) does not hold (that is $C_t=0$ or $\delta_{G_t}(u) \in N_t \cap \left( \bigcup_{i=1}^{t-1} N_i \right)$). Hence,
$$
\Prob{\delta_{H_t}(u) = v} = \Prob{\delta_{H_{t-1}}(u) = v}\left[ \Prob{C_t = 0} + \Prob{\delta_{G_t}(u) \in N_t \cap \left( \bigcup_{i=1}^{t-1} N_i } \, \wedge \, C_t = 1 \right) \right]
$$
By the induction hypothesis we have that $\Prob{\delta_{H_{t-1}}(u) = v} = 1/h_{t-1}$, and by observing that the size of $N_t \cap \left( \bigcup_{i=1}^{t-1} N_i \right)$ is $d_t + h_{t-1} - h_t$ it follows that
$$
\Prob{\delta_{H_t}(u) = v} = \frac{1}{h_{t-1}} \left( \frac{h_t - d_t}{h_t} + \frac{d_t + h_{t-1} - h_t}{d_t} \cdot \frac{d_t}{h_t} \right) = \frac{1}{h_t}
$$
\qed

\smallskip\noindent
Observe that if we look at a sequence of independent $G_{n,p}$ with $p \leqslant 1/n$ for a time-window of approximately  $1/(np)$  time steps, then every edge appears at least once in the sequence with probability at least $1/n$. The above lemma thus allows us to reduce the case $p \leqslant 1/n$ to the case $p \geqslant 1/n$. 

\begin{theorem}\label{theorem:indepunder}
Let $\mathcal{G} = \{G_t \,:\, t \in \mathbb{N}\}$ be a sequence of independent $G_{n,p}$ with $p \leqslant 1/n$
and let $s \in [n]$. The \push protocol with source $s$ over $\mathcal{G}$ completes the broadcast
 in   $\mathcal{O}(\log n / (np))$ time steps w.h.p.
\end{theorem}
\proof
Consider the sequence of random graphs $\mathcal{H} = \{H_s \,:\, s \in \mathbb{N} \}$ where $H_s$ is the union of random graphs 
\[H_s = ([n], F_s) \ \mbox{ such that } \ F_s =  E_{sT} \cup E_{sT +1} \cup  \dots \cup  E_{s T+ T-1} \ \mbox{ with }   \ T = 2/(np).  \] 
 Observe that every $H_s$ is a $G_{n,\hat{p}}$ with $\hat{p} \geqslant 1/n$. Indeed, the probability that an edge
  does not exist in $F_s$ is 
  \[ (1-p)^T \leqslant e^{-pT} = e^{-2/n}. \] Hence the probability  that the edge exists is $1 -  e^{-2/n} \geqslant 1/n$.

\noindent
Let $\tau_{\mathcal{G}}$ and $\tau_{\mathcal{H}}$ be the random variables
 indicating the completion time of the \push protocol over sequences $\mathcal{G}$ and $\mathcal{H}$ respectively. From Theorem~\ref{theorem:indepover} it follows that $\tau_{\mathcal{H}} = \mathcal{O}(\log n)$ w.h.p. and from Lemma~\ref{lemma:couplingsequences} it follows that for every $t$ it holds that 
\[
\Prob{\tau_{\mathcal{G}} \geqslant Tt} \leqslant \Prob{\tau_{\mathcal{H}} \geqslant t}.
\]
Hence, it holds that  
\[
\tau_{\mathcal{G}} = \mathcal{O}(T \log n) = \mathcal{O}\left( \frac{\log n}{np} \right) \ \mbox{ w.h.p. }
\]
\qed

\section{Edge-Markovian graphs with high dynamics} \label{sec:edgemeg}

In this section we prove that the \push protocol over an edge-Markovian graph $\mathcal{G}(n,p,q;E_0)$ with $p \geqslant 1/n$ and $q = \Omega(1)$ has completion time $\mathcal{O}(\log n)$ w.h.p.

As   observed in the Introduction, the stationary random graph is an Erd\H os-R\'enyi $G_{n,\tilde{p}}$ where $\tilde p = \frac p{p+q}$ and the mixing time of the edge Markov chain is $\Theta\left(\frac 1{p+q}\right)$. Thus, if $p$ and $q$ fall into the range defined in (\ref{parranges}), we get that the stationary random graph can be sparse and disconnected (when $p=o\left(\frac{\log n }{n}\right)$) and that the mixing time of the edge Markov chain is $O(1)$.
Thus, we can omit the term $E_0$ and assume it is random according to the stationary distribution.

The time-dependency between consecutive snapshots of the dynamic graph does not allow   us to obtain directly the \emph{increasing rate} of the number of informed nodes that we got  for  the independent-$G_{n,p}$ model. In order to get a result like Lemma~\ref{lemma:incratindip} for the edge-Markovian case, we  need in fact  a \emph{bounded-degree} condition on  the current set of informed nodes (see Definition~\ref{def:bd}) that does not apply   when the number of informed nodes is \emph{small} (i.e., smaller than $\log n$). However, in order to reach a state where at least $\log n$ nodes are informed, we can use a different ad-hoc technique that analyzes the spreading rate  yielded by  the source only. 

\begin{lemma}[The Bootstrap]\label{lemma:bootstrap}
Let $\mathcal{G} = \mathcal{G}(n,p,q)$ be an edge-Markovian graph with $p \geqslant 1/n$ and $q = \Omega(1)$, and consider the \push protocol in $\mathcal{G}$ starting with one informed node. For any positive constant $\gamma$, after $\mathcal{O}(\log n)$ time steps there are at least $\gamma \log n$ informed nodes w.h.p. 
\end{lemma}

 \proof 
We consider the message-spreading process  yielded by  the source node  only and,  instead of directly analyzing  this process  
on the edge-Markovian sequence $\{ G_t = ([n],E_t) \,:\, t \in \mathbb N \}$, we consider it  in the sequence $\{ H_t = ([n], E_{2t} \cup E_{2t+1}) \}$. Thanks to  Lemma~\ref{lemma:couplingsequences}, this is feasible since  the number of informed nodes in $H_t$ is stochastically smaller than the number of informed nodes in $G_{2t}$. We split the analysis in two cases: $p \leqslant \log n /n$ and $p \geqslant \log n / n$.

\smallskip\noindent
\underline{Case $p \geqslant \log n / n$:}
Consider an arbitrary time step $t$ during the execution of the protocol and for convenience' sake let us rename it $t=0$.
 Let $I_0$ be the set of informed nodes in that time step with $|I_0| = m \leqslant   \gamma \log n$. Consider the next two 
  time steps and let $H = ([n], E_1 \cup E_2)$ be the random graph obtained by taking the edges that are present in at least one of the two time steps.   Then  apply  the \push operation of the source node in $H$.
From Observation~\ref{obs:moresteps} (see Appendix~\ref{app:observations}), we get  that every edge has probability at least $p$ in $H$. In particular,  for every node $v$,
 the probability that $v$ is connected to the source node $s$ in $H$ is
$$
\Prob{\{s,v\} \in E_1 \cup E_2} \geqslant p\,.
$$
Let $X$ be the random variable counting the number of non-informed nodes connected to the source node in $H$, then the expectation of $X$ is
$$
\Expec{X} = \sum_{v \in [n] \setminus I_0} \Prob{\{ s,v \} \in E_1 \cup E_2} \geqslant (n-m)p \geqslant 2 \alpha np
$$
for a suitable positive constant $\alpha$. Since edges are independent, from  the Chernoff bound it follows that
$$
\Prob{X \leqslant \alpha np} \leqslant e^{-\varepsilon np}
$$
for a suitable positive constant $\varepsilon$. Hence, since $p \geqslant \log n /n$, it follows that there are at least $\alpha \log n$ nodes in  
 $[n] \setminus I_0$ that are connected to $s$ in $H$ w.h.p.
The probability that the source $s$ sends the message to one of those nodes applying  the \push operation in $H$ is
\begin{eqnarray*}
\Prob{\delta_{H}(s) \in [n] \setminus I_0} 
& \geqslant & 
\Prob{\delta_{H}(s) \in [n] \setminus I_0 \,|\, X \geqslant \alpha \log n} \Prob{X \geqslant \alpha \log n} \\
& \geqslant &
\frac{\alpha \log n}{m + \alpha \log n} \Prob{X \geqslant \alpha \log n} 
\geqslant \lambda
\end{eqnarray*}
for a suitable positive constant $\lambda$.

\noindent
From Lemma~\ref{lemma:couplingsequences},   the probability that the actual number $m_{2}$ of informed nodes after two time steps is smaller than $m_0 +1$ is  
at most as large as the probability that the source node informs a new neighbor in $H$; i.e.,
$$
\Prob{m_{2} = m_0}\leqslant \Prob{\delta_{H}(s) \notin [n] \setminus I_0} \leqslant 1-\lambda\,.
$$ 

\noindent
Thus for every time step $t$ during the bootstrap, if $p \geqslant \log n / n$,  after two time steps there is at least one new informed node with probability at least $\lambda$; i.e.,
$$
\Prob{m_{t+2} \geqslant m_{t} + 1} \geqslant \lambda\,.
$$
Hence, after $(4 \gamma / \lambda) \log n$ time steps, there are at least $\gamma \log n$ informed nodes w.h.p.

\smallskip\noindent
\underline{Case $p \leqslant \log n /n$:}
In order to analyze the bootstrap phase on the sequence $\{ H_t = ([n], E_{2t} \cup E_{2t+1}) \}$, we first condition on the event $\overline{F}$ that in the first $T=(4\gamma/\lambda)\log n$ time steps it never happens that a new edge appears between the source node and a node that is already informed. Formally, $\overline{F}$ is the complementary event of $F := \cup_{t=1}^T F_t$ where $F_t$ denotes the event ``In $H_{t+1}$ at least one edge will appear between the source node and a previously informed node''. 
As we will  see below, we have $\Prob{F} = \mathcal{O}(\log^3n / n)$ and 
$\Prob{|I_T| \leqslant \gamma \log n \,|\, \overline{F}} \leq n^{-\varepsilon}$  for a suitable positive constant $\varepsilon$.
 
\noindent
 Observe that if an edge does not exist in $H_t$ then it will appear in $H_{t+1}$
 with probability $1 - (1-p)^2$.    Since $p \leqslant \log n /n \leqslant 1/4$, by applying the standard inequalities 
  $e^{-2x}  \leqslant 1 - x \ \leqslant  e^{-x}$,   for any $ 0  \leqslant  x    \leqslant    \frac 12$,
  we get 
$
2p \leqslant 1 - (1-p)^2 \leqslant 4 p$.
For $F_t$ as defined above we have
\begin{equation}\label{eq:effet}
\Prob{F_t} \leqslant 4p |I_t| \leqslant 4 \gamma \frac{\log^2 n}{n}\,,
\end{equation}
where in the last inequality we used the facts that  $p \leqslant \log n / n$ and  that, during the bootstrap,  $|I_t| \leqslant \gamma \log n$.

\noindent
Now consider the two following events: $S_1^t$ is the event ``The source informs a new node  in $H_{t+1}$'' and $S_2^t$ is the event ``The number of edges between the source node and the set of informed nodes decreases \ in $H_{t+1}$''; i.e.,  
$
S_1^{t} = \{|I_{t+1}| = |I_t|+1 \}$    and $S_2^t = \{ \grad_{t+1}(s,I_{t+1}) \leqslant \grad_t(s, I_{t}) - 1 \}$.
 Now we show that, at every time step, at least one of the  two events above holds with constant probability if event $F_t$ does not hold. Indeed, in that case, if the number of informed nodes connected to the source node is zero, then if some non-informed node will be connected to the source node at the following time step we will have at least a new informed node (event $S_1^t$) and this  happens with constant probability. If there is at least one informed node connected to the source, then if one of those edges will disappear then $\grad(s, I_t)$ will decrease (event $S_2^t$). More formally, if $\grad_t(s,I_t) = 0$ we have that
\[
\Prob{S_1^t \,|\, \overline{F_t}} 
\geqslant 1 - (1-2p)^{n-|I_t|}
\geqslant 1 - e^{-2p (n-|I_t|)}
\geqslant 1 - e^{-(2/n) (n-|I_t|)}
\geqslant 1 - e^{-1}\,.
\]
If $\grad_t(s,I_t) \geqslant 1$,  we get 
$ \Prob{S_2^t \,|\, \overline{F_t}} \geqslant q$.
Hence for $\lambda = \min \{q, \, 1- e^{-1} \}$,  we have that
\begin{equation}\label{eq:boundprobotheffe}
\Prob{S_1^t \vee S_2^t \,|\, \overline{F_t}} \ \geqslant \ \lambda\,.
\end{equation}
If we define  $T = (4 \gamma / \lambda) \log n$ then  we can  show that after $T$ time steps there are at least $\gamma \log n$ informed nodes w.h.p. Indeed, let  $X_1$ and $X_2$ be the random variables indicating the number of time steps that events $S_1$ and $S_2$ hold, respectively.  
 Remind  that its complement $\overline{F}$ is the event ``In the first $T$ time steps it never happens that a new edge appears between the source node and a node that is already informed''. Since $T = \mathcal{O}(\log n)$, from Eq.~\ref{eq:effet} it follows that  
 $\Prob{F} =\mathcal{O}(\log^3n / n)$.
 Moreover, observe that if event $\overline{F}$ holds then $X_1 \geqslant X_2$. Indeed, if no edge between the source   and  any previously informed node appears, then, when an edge between the source node and an informed node disappears (event of $S_2$ type),  the source must have previously informed that node ($S_1$ event).
Thus the probability that the bootstrap is not completed at time $T$ is
\[
\Prob{|I_T| \leqslant \gamma \log n}  \ 
 \leqslant \  \Prob{X_1 \leqslant \gamma \log n \,|\, \overline{F}} + \Prob{F} \  
 \leqslant  \  \Prob{X_1 + X_2 \leqslant 2 \gamma \log n \,|\, \overline{F}} + \Prob{F}. \]
 Since from Eq.~\ref{eq:boundprobotheffe} we have that, at every time step, the 
  event $S_1 \vee S_2$ holds with probability at least $\lambda$, then 
$
\Prob{X_1 + X_2 \leqslant 2 \gamma \log n \,|\, \overline{F}}
$
is smaller than the probability that in a sequence of $T = (4 \gamma / \lambda) \log n$ independent coin tosses, each one giving \texttt{head} with probability $\lambda$, we see less than $2 \gamma \log n$ \texttt{heads}: this is smaller than 
$n^{-\varepsilon}$ for a suitable positive constant $\varepsilon$.
\qed


\medskip\noindent
We can now start the second part of our analysis where the \push operation of all informed nodes (forming the subset $I$) will be considered and, thanks to the bootstrap, we can assume that   $|I| = \Omega(\log n)$.

\noindent
As mentioned at the beginning of the section, we need to introduce the concept of \emph{bounded-degree state} $(E,I)$ of the Markovian process describing the information-spreading process over the dynamic graph, where $E$ is the set of edges and $I$ is the set of informed nodes.

\begin{definition}[Bounded-Degree  State] \label{def:bd}
A state $(E,I)$ such that $|E(I)| \leqslant (8/q) n \tilde{p} |I|$ (with  $\tilde{p}=\frac{p}{p+q}$ the stationary edge probability) will be called a \emph{bounded-degree} state.
\end{definition}

\smallskip
\noindent
In the next lemma we show that, if $I$ is the set of informed nodes with $|I| \geqslant \log n$, if in the starting random graph $G_0$ every edge exists with probability approximately $(1\pm \varepsilon) p$, and if it evolves according to the edge-Markovian model and the informed nodes perform the \push protocol, then for a long sequence of time steps the random process is in a bounded-degree state. We will use this property in Theorem~\ref{theorem:edge} by observing that, for every initial state, after $\mathcal{O}(\log n)$ time steps an edge-Markovian graph with $p \geqslant 1/n$ and $q \in \Omega(1)$  is in a state where every edge $\{u,v\}$ exists with probability $p_{\{u,v\}} \in \left[ (1-\varepsilon) \tilde p,\, (1+\varepsilon) \tilde p \right]$.
 
\begin{lemma}\label{lemma:bdsequence}
 Let $\mathcal{G} = \mathcal{G}(n,p,q,E_0)$ be an edge-Markovian graph starting with $G_0$ and consider the \push protocol in $\mathcal{G}$ where $I_0$ is the set of informed nodes at time $t=0$. Then, for any constant $c>0$,  for a sequence of $c\log n$ time steps every state is a bounded-degree one w.h.p.
\end{lemma}
\proof
Let us fix $c = 8/q$ as in Definition~\ref{def:bd}. We show that $(E_0,I_0)$ is a bounded-degree state w.h.p. and that if $(E_t,I_t)$ is  a bounded-degree state, then $(E_{t+1},I_{t+1})$ is a bounded-degree state as well w.h.p.
Let us name $X_t = | E_t(I_t) |$. The expected size of $E_0(I_0)$ is
 
$$
\Expec{X_0} \leqslant \left[ \binom{|I_0|}{2} + |I_0| (n-|I_0|)  \right] (1+\varepsilon) \tilde p \leqslant (1 + \varepsilon) n \tilde p |I_0|\,.
$$
Since edges are independent, $c \geqslant 8$, and $n \tilde p |I_0| = \Omega(\log n)$, from  Chernoff bound it follows that $|E_0(I_0)| \leqslant c n \tilde p |I_0|$ w.h.p.
Now let $t \geqslant 0$ and assume that $X_t \leqslant c n \tilde p |I_0|$. Observe that the size of $E_{t+1}(I_{t+1})$ satisfies 

\begin{equation}\label{eq:split}
X_{t+1} \ = \ |E_{t+1}(I_{t}) | + |E_{t+1}(\hat I_{t+1}, [n] \setminus I_t)|\,,
\end{equation}
where $\hat I_{t+1}:=I_{t+1} \setminus I_t$.
As for the first addend, we have that
\begin{eqnarray*}
\Expec{|E_{t+1}(I_{t})| \;|\; X_t} & = & (1-q) X_t + p \left[ \binom{|I_t|}{2} + |I_t| (n-|I_t|) - X_t \right] \\
& = & \left(1- (p+q) \right) X_t + p \left[ \binom{|I_t|}{2} + |I_t| (n-|I_t|) \right]
\end{eqnarray*}
because all the $X_t$ edges existing at time $t$ are still there at time $t+1$ with probability $1-q$ and all the edges that do not exist at time $t$ appear with probability $p$. Since
$p = \tilde p(p+q) \leqslant 2 \tilde p$, if $p + q \geqslant 1$ then
$$
\Expec{|E_{t+1}(I_{t})|} \leqslant 2 n \tilde p |I_t| \leqslant \frac{q}{4} c n \tilde p |I_t|\,,
$$
regardless of the value of $X_t$. 
 If instead   $p + q \leqslant 1$ then, if $X_t \leqslant c n \tilde p |I_t|$ we have that 
 
\begin{eqnarray}\label{eq:firstaddend}
\Expec{|E_{t+1}(I_{t})| \;|\; X_t \leqslant c n \tilde p |I_t|}
& \leqslant & \left( 1-p-q \right) c n \tilde p |I_t| +  n p |I_t| \nonumber \\
& = & c n \tilde p |I_t| \left( 1-p-q + \frac{(p+q)}{c} \right) \nonumber \\
& \leqslant & \left( 1 - \frac{q}{2} \right) c n \tilde p |I_t|\,,
\end{eqnarray}
where in the last inequality we used that $p \geqslant 0$ and $(p+q)/c \leqslant q/2$.

\noindent
As for the second addend, we observe that every pair $e = \{u,v\}$ with $u \in \hat{I}_{t+1}$, $v \in [n]\setminus I_t$, and $u \neq v$ exists in $E_{t+1}(\hat I_{t+1}, [n] \setminus I_t)$ with probability $p_e \in \left[ (1-\varepsilon) \tilde p,\, (1+\varepsilon) \tilde p \right]$ since it has never been observed before time $t+1$. Hence
\begin{equation}\label{eq:secondaddend}
\Expec{|E_{t+1}(\hat I_{t+1}, [n] \setminus I_t)|}
\leqslant |\hat{I}_{t+1}| (n - |I_t|)(1+\varepsilon) \tilde p  
\leqslant \frac{q}{4} c n \tilde p |I_t|\,.
\end{equation}
By (\ref{eq:firstaddend}) and (\ref{eq:secondaddend}) in (\ref{eq:split}) we get
$$
\Expec{X_{t+1} \;|\; X_t \leqslant c n \tilde p |I_t|}
\leqslant \left(1 - \frac{q}{4} \right)c n \tilde p |I_t| \leqslant \left(1 - \frac{q}{4} \right)c n \tilde p |I_{t+1}|\,.
$$
Since edges are independent, $q = \Omega(1)$, and $n \tilde p |I_{t+1}| = \Omega(\log n)$, from Chernoff bound it follows that $X_{t+1} \leqslant c n \tilde p |I_{t+1}|$ w.h.p.
\qed

\smallskip
\noindent
Now we can bound  the \emph{increasing rate} of the number of informed nodes in an edge-Markovian graph. The proof of the following lemma combines the analysis adopted in the proof of Lemma~\ref{lemma:incratindip} with some further ingredients required to manage the time-dependency of the edge-Markovian model.

\begin{lemma}[The increasing rate of new informed nodes]
\label{lemma:incratedge}
Let $(E,I)$ be a bounded-degree state and let $X$ be the random variable counting the number of non-informed nodes that get informed after two steps of the \push operation in the edge-Markovian graph model. It holds that
$
\Prob{X \geqslant \varepsilon \cdot \min\{ |I|, \, n-|I| \}} \geqslant \lambda$,
where $\varepsilon$ and $\lambda$ are positive constants.
\end{lemma}
\proof
 Let $G_0 = ([n],E_0)$ be the current graph and let $G_1 = ([n],E_1)$ and $G_2 = ([n],E_2)$ be the next two random graphs obtained according to the edge-Markovian process starting from  $G_0$. Let $H = ([n],E_H)$ be such that $E_H = E_1 \cup E_2$ and let $\hat{H}$ be the $(I, 3cn \tilde{p})$-modified version of $H$ according to Definition~\ref{def:modifiedgraph}, where $c$ is a sufficiently large constant (it will be clear from what follows that it is sufficient to have  $c \geqslant 32/q$). From Lemmas~\ref{lemma:virtualnodes} and~\ref{lemma:couplingsequences}, we have that the number of informed nodes in $\hat{H}$ is stochastically smaller than the number of informed nodes in $G_{2}$. In what follows we evaluate the number of new informed nodes in $\hat{H}$ and we show that with positive constant probability it is at least a constant fraction of $\min\{ |I|, \, n-|I| \}$.

Let $I_A$ be the set of informed nodes that have degree at most $c n \tilde{p}$, i.e., 
\[
I_A = \{ u \in I \,:\, \grad_{G_0}(u) \leqslant c n \tilde{p} \}\,.
\] 
In what follows, $I_A$ will denote the set of \emph{active} informed nodes. Observe that
$$
\sum_{u \in I} \grad_{G_0}(u) \leqslant 2 |E(I)|.
$$
Since $(E,I)$ is a bounded-degree state, we have $2 |E(I)| \leqslant (16/q) n\tilde{p}|I|$. Thus, if $c \geqslant 32/q$ then we have that $|I_A| \geqslant |I|/2$.

\noindent
Consider an active informed node $u \in I_A$ and let $v \in [n] \setminus I$ be a non-informed one. The probability that node $u$ selects node $v$ in $\hat{H}$ according to the \push protocol is
\begin{eqnarray}\label{eq:uvhatH}
\Prob{\delta_{\hat{H}}(u) = v} & = & \Prob{\delta_{\hat{H}}(u) = v \;|\; \{ u,v \} \in E_H, \, \grad_H(u) \leqslant 3 c n \tilde{p} } \cdot \nonumber \\ 
& & \phantom{some space} \cdot \Prob{\grad_H(u) \leqslant 3 c n \tilde{p} \; | \; \{ u,v \} \in E_H} \Prob{\{ u,v \} \in E_H}\,.
\end{eqnarray}
Indeed, by the definition of $\hat{H}$, $u$ cannot select $v$ in $\hat{H}$ if the edge $\{ u,v \}$ does not exist in $H$ or if the degree of $u$ in $H$ is larger than $3 c n \tilde{p}$.

\noindent
Now observe that 
\begin{equation}\label{eq:uvhatHc1}
\Prob{\delta_{\hat{H}}(u) = v \,|\, \{ u,v \} \in E_H, \, \grad_H(u) \leqslant 3 c n \tilde{p} } \geqslant 1/(6cn\tilde{p}) \,.
\end{equation}
Indeed, node $u$ has $3cn\tilde{p}$ virtual neighbors in $\hat{H}$ plus up to $3 c n \tilde{p}$ non-informed neighbors.
As for $\Prob{\{ u,v \} \in E_H}$, from Observation~\ref{obs:moresteps} (see Appendix~\ref{app:observations}), it follows that
\begin{equation}\label{eq:uvhatHc2}
\Prob{\{ u,v \} \in E_H} \geqslant p = \tilde{p} (p+q) \geqslant q \cdot \tilde{p}\,.
\end{equation}
We now show that $\Prob{\grad_H(u) \leqslant 3 c n \tilde{p} \; | \; \{ u,v \} \in E_H}$ is larger than a positive constant. Observe that we can write
$$
\grad_H(u) = \sum_{w \in [n]\setminus\{u\}} X_w\,,
$$
where $X_w$ is the indicator random variable of the event $\{ u,w \} \in E_H$. Thus,
\begin{equation}\label{eq:expecdeg1}
\Expec{\grad_{H}(u) \,|\, \{u,v\} \in E_H} = \sum_{w \in [n] \setminus \{u\}} \Prob{X_w = 1 \,|\, \{u,v\} \in E_H}\,.
\end{equation}
Now observe that, for $w \neq v$, $\Prob{X_w = 1\,|\, \{u,v\} \in E_H} = \Prob{X_w = 1}$ and it can have two values, depending on whether or not edge $\{u,w\}$ existed in $G_0$,
\begin{eqnarray*}
\Prob{X_w = 1 \,|\, \{u,w\} \notin E_0} & = & p + (1-p)p\,, \\[2mm]
\Prob{X_w = 1 \,|\, \{u,w\} \in E_0} & = & 1-q + qp\,.
\end{eqnarray*} 
Hence, if we split the sum in~(\ref{eq:expecdeg1}) in the $w$'s that were neighbors of $u$ in $E_0$ and those that were not, we get 

\begin{eqnarray*}
\Expec{\grad_{H}(u) \,|\, \{ u,v \} \in E_H} & \leqslant & 1 + (1-q + qp) \grad_{G_0}(u) + (n - \grad_{G_0}(u)) (p + (1-p)p) \\
& \leqslant & 1 + \grad_{G_0}(u) + (n - \grad_{G_0}(u)) 2 p \\
& \leqslant & c n \tilde{p} + 3 n p \\
& \leqslant & 2c n \tilde{p}\,,
\end{eqnarray*}

\noindent
where, from the first line to the second one we used that $p + (1-p)p \leqslant 2p$ and $1-q + qp \leqslant 1$, from the second to the third line we used that $1 \leqslant np$ and that $\grad_{G_0}(u) \leqslant c n \tilde{p}$, because $u \in I_A$, and from the third line to the fourth one we used that $p = (p+q)\tilde{p} \leqslant 2 \tilde{p}$ and $c \geqslant 6$.  From Markov's inequality it thus follows that
\begin{equation}\label{eq:uvhatHc3}
\Prob{\grad_{H}(u) \geqslant 3 n \tilde{p} \,|\, \{u,v\} \in E_H} \leqslant 2/3\,.
\end{equation}
By combining (\ref{eq:uvhatHc1}), (\ref{eq:uvhatHc2}), and (\ref{eq:uvhatHc3}) in (\ref{eq:uvhatH}) we get
$$
\Prob{\delta_{\hat{H}}(u) = v} \geqslant \frac{\alpha}{n}
$$
for a suitable positive constant $\alpha$.

\noindent
Since the events $\{ \delta_{\hat{H}}(u) \neq v \,:\, u \in I_A \}$ are independent, the probability that node $v$ is not informed in $\hat{H}$ is
\[
\Prob{\bigcap_{u \in I_A} \delta_{\hat{H}}(u) \neq v} 
\leqslant \left( 1 - \alpha / n \right)^{|I_A|}
\leqslant e^{- \alpha |I_A| / n}
\leqslant e^{- (\alpha/2) |I| / n}\,.
\]
\smallskip\noindent
Let $X$ be the random variable counting the number of new informed nodes in $\hat{H}$. The expectation of $X$ is thus
$$
\Expec{X} \geqslant (n-|I|)\left( 1 - e^{- (\alpha/2) |I| / n} \right) \geqslant (\alpha / 4) (n-|I|)|I| / n\,.
$$
Hence we have that
$$
\Expec{X} \geqslant
\left\{
\begin{array}{cl}
(\alpha / 8) |I| & \quad \mbox{ if } |I| \leqslant n/2\,, \\[2mm]
(\alpha / 8) (n-|I|) & \quad \mbox{ if } |I| \geqslant n/2\,.
\end{array}
\right.
$$
Since $X \leqslant \min \{ |I|,\, n-|I| \}$ the thesis then follows from Observation~\ref{obs:expecttoprob} (see Appendix~\ref{app:observations}).
\qed

\smallskip
\noindent
Now we can prove that in $\mathcal{O}(\log n)$ time steps the \push protocol informs all nodes  in an edge-Markovian graph, w.h.p.

\begin{theorem}\label{theorem:edge}
Let $\mathcal{G} = \mathcal{G}(n,p,q,E_0)$ be an edge-Markovian graph with $p \geqslant 1/n$ and $q = \Omega(1)$ and let $s \in [n]$ be a node. The \push protocol with source $s$ completes the broadcast over $\mathcal{G}$ in $\mathcal{O}(\log n)$ time steps w.h.p.
\end{theorem}
\proof
Lemma~\ref{lemma:bootstrap} implies that after $\mathcal{O}(\log n)$ time steps there are $\Omega(\log n)$ informed nodes w.h.p. From Observation~\ref{obs:moresteps} (see Appendix~\ref{app:observations}) and Lemma~\ref{lemma:bdsequence},  it follows that, after further $\mathcal{O}(\log n)$ time steps, the edge-Markovian graph reaches    a bounded-degree state and remains so  for further $\Omega(\log n)$ time steps.
Let us rename $t=0$ the time step where there are $\Omega(\log n)$ informed nodes and every edge $e \in \binom{[n]}{2}$ exists with probability
 $p_e \in [(1-\varepsilon) \tilde p \,,\; (1+\varepsilon)\tilde p]$. 
We again abbreviate   $m_{t}:=|I_{t}|$.
Observe that if recurrence $m_{2(t+1)} \geqslant (1+\varepsilon)m_{2t}$ holds $\log n / \log (1+\varepsilon)$ times, then there are $n/2$ informed nodes. Let us thus name
$
T = \frac{2}{\lambda} \frac{\log n}{\log (1+\varepsilon)}
$.
If at time $2T$ there are less than $n/2$ informed nodes, then recurrence $m_{2(t+1)} \geqslant (1+\varepsilon)m_{2t}$ held less than $\lambda T / 2$ times. Since, at each time step,  the recurrence  holds with probability at least $\lambda$ (there are less than $n/2$ informed nodes and the state is a bounded-degree one w.h.p.), the above probability is 
at most as large as the probability that in a sequence of $T$ independent coin tosses, each one giving \texttt{head} with probability $\lambda$, we see less than 
$(\lambda/2)T$  \texttt{heads} (see, e.g., Lemma 3.1 in \cite{ABKU99}).
By the Chernoff bound such a probability is smaller than $e^{-\gamma \lambda T}$, for a suitable positive constant $\gamma$. Since $\gamma$ and $\lambda$ are constants and $T = \Theta(\log n)$ we have that
\begin{equation}\label{eq:edgen2}
\Prob{m_{2T} \leqslant n/2} \leqslant n^{-\delta}
\end{equation}
for a suitable positive constant $\delta$.
When $m_t$ is larger than $n/2$ and the edge-Markovian graph is in a bounded-degree state, from Lemma~\ref{lemma:incratedge} it follows that recurrence $n - m_{t+1} \leqslant (1-\varepsilon) (n - m_t)$ holds with probability at least $\lambda$. If this  recurrence holds $\log n / \log\left( 1/(1-\varepsilon) \right)$ times then the number of informed nodes cannot be smaller than $n$. Hence, if we name $\tilde{T} := (2/\lambda) \log n / \log\left( 1/(1-\varepsilon) \right)$, with the same argument we used to get (\ref{eq:edgen2}), we obtain that after $2T+2\tilde{T}$ time steps all nodes are informed w.h.p.
\qed


\section{Conclusions} \label{sec:conc}

In this paper we studied the \push protocol over edge-MEGs. We first analyzed the independent $G_{n,p}$ case (i.e. the edge-MEG with $q = 1-p$) and we showed that the completion time is $\mathcal{O}(\log n / n\hat{p})$ w.h.p., where $\hat{p} = \min\{p, \, 1/n\}$. Then we studied the general edge-MEG model  with $p \geqslant 1/n$ and $q = \Omega(1)$ and we showed that the completion time is logarithmic.  This bound is obviously tight because the \push protocol cannot inform $n$ nodes in less than $\log_2 n$ time steps. 

Our results can be extended to the case of ``more static'' sparse dynamic graphs. Indeed, we can provide a logarithmic bound on the completion time of  the \push protocol over the $\sG(n,p,q)$ model even for  $p = \Theta(1/n)$ and for $q = o(1)$. The proof of the following result combines some new  coupling arguments with a previous analysis of the \push protocol    for static random graphs given  in  \cite{FPRU90} (a sketch of the proof is given in Appendix~\ref{app:proofofthm:slow}). 

\begin{theorem} \label{thm:slow}
 Let $p = \frac {d}{n}$ for some  absolute constant $d \in \mathbb N$ 
and let     $q=q(n)$ be such that $q(n) =o(1)$.
The \push protocol over  edge-Markovian graphs in $\sG(n,p,q)$ completes in $O(\log n)$ time, w.h.p.
\end{theorem}

We believe that the most challenging question is to analyze rumor spreading over  more general classes of evolving graphs where edges may be not independent: for instance, it would be interesting to analyze the \push protocol over geometric models of mobile networks \cite{CMPS09,Jacquet}.


\bibliographystyle{plain}
\bibliography{dynpush}

\newpage
\appendix
\centerline{\Huge\bf Appendix}
\bigskip


\section{Sketch of proof for Theorem~\ref{thm:slow}}
\label{app:proofofthm:slow}

The proof makes use of the following previous result.

\begin{lemma}[Theorem 12 in  \cite{FPRU90}] \label{thm:staticrnd}
For any $\varepsilon >0$,  consider  an \ErdRen random graph $\sG(n,p)$ with $p\geq (1+\varepsilon)\frac{\log n}{n}$.
Then, the  \push protocol has w.h.p. completion time    $\Theta(\log n)$.
\end{lemma}

\noindent
We start by giving    an equivalent formulation of the edge-Markovian model.
 Let $e=\{u,v\}$ be a pair of nodes (unordered) and 
$t\in\mathbb{N}$. We define  two families of Bernoulli random variables $\{U_{e, t}\}$ and $\{V_{e, t}\}$
with parameters $\hat{p}$ and $\hat{q}$ respectively. At each time step $t$, we first set edge $e$
to {\em empty} if $V_{e, t}=1$ and leave it unchanged if $V_{e, t}=0$; then we set edge $e$
to {\em full} if $U_{e, t}=1$ and leave it unchanged if $U_{e, t}=0$.

\noindent
It is easy to verify  that this process is   equivalent to the $\sG(n,p,q)$ process by taking
$p=\hat{p}$ and $q=\hat{q}(1-\hat{p})$, as long as $1-p=\Theta(1)$.

\noindent
  It is also useful to consider the following partial order on node \emph{configurations} $(I,[n]\setminus I)$, where $I$ is  the subset of the informed nodes. 
We say that configuration ${\cal C}$ is \emph{below} configuration ${\cal C'}$ if every informed
node of ${\cal C}$  is also an informed node of ${\cal C'}$.

\noindent
In order to   prove the theorem,   we need to analyze   some   ranges for $q = q(n)$ separately.
 
\medskip
\noindent
{\large {\bf -  $q(n)=o(1/\log n)$}.}
Under this condition,  
the stationary graph   is w.h.p. fully connected with $\tilde{p}=\omega(\frac{\log n}{n})$.
Moreover w.h.p. the degree of every node is larger than 
$\alpha n q(n)$ for some (small) positive constant $\alpha$. The key observation here is to observe that
the death rates are so small that a static approximation will suffice. 
We make this idea  more formal  by introducing another coupling that    requires this time to look into the future.
Let's look at the evolution of the edges  for $k \log n$ steps, where $k$ is a (sufficiently) large constant and mark
all the edges  that will die during that time period.
We   now modify the dynamics as follows: whenever a marked edge  is selected by the \push  to transmit
the message,  then the transmission does not take place. This process  is clearly below the one we are
considering, under the partial order introduced above. Thus the completion time $T$ of  the new process
is larger than that of   the original one. 

\noindent
Observe that, for  each node, the probability to ever be denied the use of an edge,   within the time window 
under consideration, is only $o(1)$. This makes the dynamics only negligibly slower and
therefore the completion time  $T$ will be only  a constant-factor  larger than that in the process  with no deaths.
We can thus apply Lemma \ref{thm:staticrnd} and get the thesis.

\medskip
\noindent
{\large {\bf - $q(n)$ from $O(1/ \log n)$ to $o(1)$.}
} Under this condition, the stationary graph has edge probability  $\tilde{p}=\frac{1}{nq}$ and only $o(n)$ nodes
 do not belong to the giant component. 
Moreover the average degree is $\Theta(1/q)$ and, by a standard application of Chernoff's bound,
   the probability that a node has degree between $\alpha / q(n)$
and $\beta / q(n)$ is bounded by $\exp(-\frac{M}{q_n})$ for some real  $M$ depending on $\alpha$ and
$\beta$  but not on $n$. The analysis of the \push protocol is organized in stages.

 \noindent
{\em - Stage 0:} If the source  node does not belong to the giant component, we only need to wait 
$O(1/q(n))$ steps for the message  to infect   one node of  the giant component. 
If the source  node belongs to the giant component, this stage can be skipped.

\noindent
{\em - Stage 1:}  Let   $m_t = |I_t|$ be the number of informed nodes at time $t$. This stage concerns
the process while $m_t$ is  in the range $1 \leqslant m_t \leqslant \gamma n$, for some absolute constant $\gamma >0$. 
We will consider a modification of the process  so that a node  is only allowed to
transmit the message for  $k$ times, where $k$ will be fixed later. Clearly,  the modified  process  is below the original one.
Let $A$ be the bad event  ``an informed node   is selected by the \push  to receive the source message''.
Then observe that  
\[ \Prob{A} \ \leqslant \  kq(n)+\frac{m_{t+1}}{n} \ \leqslant \     \gamma' , \ \mbox{ for some constant } \gamma' \]
This implies
\[
\Expe{m_{t+1}\,|\,m_t} \geqslant \ m_t+ (1- \Prob{A})(m_t-m_{t-k}) \ \geqslant   \ 
m_t+ (1- \gamma')(m_t-m_{t-k})
\]
Taking the expectation and setting $\Expe{m_t}=\mu_t$, we have
\[
\mu_{t+1} \ \geqslant \  (2- \gamma')\mu_t +(1- \gamma')\mu_{t-k}
\]

\noindent
Now, we can choose $\gamma \in (0,1)$ (thus $\gamma'$) and $k\in \mathbb{N}$   so that the equation 
\[ z^{k+1} \ - \ (2- \gamma')z^k \ - \ (1- \gamma') \] has one root larger than 1.
This ensures exponential growth of $\mu_t$ 
and thus completion time of Stage 1 in $O(\log n)$ steps.
Observe that the above bound holds w.h.p. Indeed,
let  $\delta$ be  the largest root of the above indicial equation. 
Since $m_t$ is a Markov chain, the events 
\[ \{m_{t+1}>\Expe{m_{t+1}\,|\,m_t}\} \] are independent 
for different $t$'s. Moreover we have the deterministic bounds 
\[   m_t  \ \leqslant  \ m_{t+1} \leqslant  2 \ m_t \]
From this,  we get that (e.g from the Paley-Zygmund inequality)
\[ \Prob{m_{t+1} > \Expe{m_{t+1}\,|\,m_t}}\geqslant \eta>0  \]
 By a standard application of Chernoff's Bound, for any  integer $c$, we can fix a 
 suitable constant $D$ such that, 
after  $t \geqslant D\log n$ steps , we get $\Prob{m_t>\delta^{\eta t}} \geqslant
 1-\frac{1}{t^c}$.

\noindent
{\emph - Stage 2:} After Stage 1, by waiting $O(k/q(n))$ steps we can ensure that, w.h.p.,   for every  node $v$, 
 an arbitrarily-large constant fraction  of the $v$-edges  
  will be new, i.e. they were not in existence at the end of Stage 1. This is equivalent to randomizing
the informed nodes.

\noindent
{\emph - Stage 3:} We now consider a  node $v$ and estimate the probability that $v$ has not received
information after $D\log n$ further steps. We call a vertex {\em good} if it has degree between 
$\alpha / q(n)$ and $\beta /q(n)$, otherwise we call it {\em bad}.
First observe that for arbitrarily  small $\varepsilon >0$ and $n$ large enough, it holds 
\[
e^{- \frac{M-\varepsilon}{q(n)}} \ < \ D\log n\, e^{-\frac{M}{q(n)}} \ < \ e^{-\frac{M}{q(n)}}
\]
So that the probability that a node  is ever bad in a time interval of  length $D\log n$
is bounded by $e^{\frac{M-\varepsilon}{q(n)}}$. Let $v$ be good for all the time.
The probability that the source message is not transmitted to $v$ in a given step is bounded above by
\[
\left(1-\frac{q(n)}{\beta}\right)^{\gamma' \frac{\alpha}{q(n)}} \ \simeq \ e^{-\gamma'\frac{\alpha}{\beta}}
\]
Now, after $\frac{4\beta}{\gamma\alpha}\log n$ steps,  the probability the $v$ has not received
the message is bounded by $n^{-4}$. So the probability that there is a good vertex which has not yet been 
informed is bounded by $n^{-2}$.

\noindent
{\em Stage 4:}   We are now left with at most  $O(n\,e^{\frac{M-\varepsilon}{q(n)}})$   non-informed nodes.
In order to  show that they have actually been informed during Stage 3, we need to look more carefully  at how the degree
of a given node  evolves in time. This is a Markov chain on $[0, \ldots, n]$ with stationary  measure $\mu$ which
is binomial with parameters $(n, \frac{1}{nq(n)})$.
 As we observed before, it holds that 
   \[ \mu ([\alpha q_n^{-1}, \beta q_n^{-1}]) \
    \geqslant \  1- e^{\left(-\frac{M}{q(n)}\right)} \] By taking $D$ large
 enough,  we get that  the chain  will spend a positive fraction of the time in $[\alpha / q(n), \beta q(n)]$ with
 probability at least $1-\frac{1}{n^4}$. We then get that the probability that there is a pair  of nodes  which are
   both bad for a positive fraction of the time is bounded by $n^{-2}$.
By restricting information transmission to pairs  of  good nodes, we can again use the analysis of    Stage 3.
 
\medskip
 \noindent
{\large {\bf - $q(n) \ = \ O(1/ \log n)$. }}
This case is similar to previous one, but it is easier, so it will be omitted.

\section{A few observations}
\label{app:observations}

\begin{obs}\label{obs:moresteps}
Consider the general two state Markov chain
$$
\left(
\begin{array}{c|cc}
 & 0 & 1 \\
\hline
0& 1-p & p \\
1 & q & 1-q
\end{array}
\right)
$$
Then 
\begin{itemize}
\item For every initial state $x \in \{0,1\}$, the probability that the chain is is state $1$ in at least one of the first two time steps is
$$
\Prob{X_2 = 1 \mbox{ or } X_1 = 1 \,|\, X_0 = x} \geqslant p
$$
\item Let $p_t = \Prob{X_t = 1}$ be the probability that the chain is in state $1$ at time $t$. Then
$$
p_t = \frac{p}{p+q} + \left( p_0 - \frac{p}{p+q} \right) (1-p-q)^t 
$$
\end{itemize}

\end{obs}

\begin{obs}\label{obs:expecttoprob}
Let $X$ be a random variable taking values between $0$ and $m$, for some positive real $m$. If $\Expec{X} \geqslant \lambda m$ for some $0 \leqslant \lambda \leqslant 1$, then 
$$
\Prob{X \geqslant \frac{\lambda}{2} m} \geqslant \lambda / 2
$$
\end{obs}

\end{document}